\documentclass[runningheads]{llncs}
\usepackage[T1]{fontenc}
\usepackage{graphicx}
\usepackage{amsmath,amssymb, amsfonts, bm, csquotes}
\usepackage{physics, algorithm, subcaption}
\usepackage[noend]{algpseudocode}
\algnewcommand{\LineComment}[1]{\State \(\triangleright\) #1}

\usepackage[colorinlistoftodos]{todonotes}

\begin{document}
\title{Benchmarking ORCA PT-1 Boson Sampler in Simulation}
%
%
\author{Jessica Park\inst{1}\orcidID{0000-0001-9001-9036} \and
Susan Stepney\inst{1}\orcidID{0000-0003-3146-5401} \and
Irene D'Amico\inst{2}\orcidID{0000-0002-4794-1348}}
\authorrunning{J. Park et al.}
%
\institute{Department of Computer Science, University of York, UK \newline
\email{jlp567@york.ac.uk} \and
Department of Physics, Univeristy of York, UK}
\maketitle              

\begin{abstract}
Boson Sampling,  a non-universal computing paradigm, has resulted in impressive claims of quantum supremacy. 
ORCA Computing have developed a time-bin interferometer (TBI) that claims to use the principles of boson sampling to solve a number of computational problems including optimisation and generative adversarial networks \cite{Bradler2021-xa}. 
We solve a dominating set problem with a surveillance use case on the ORCA TBI simulator to benchmark the use of these devices against classical algorithms. Simulation has been used to consider the optimal performance of the computing paradigm without having to factor in noise, errors and scaling limitations. 
We show that the ORCA TBI is capable of solving moderately sized ($n<250$) dominating set problems with comparable success to linear programming and greedy methods. 
Wall clock timing shows that the simulator has worse scaling than the classical methods, but this is unlikely to carry over to the physical device where the outputs are measured rather than calculated.

\end{abstract}
\section{Introduction}

Boson sampling is a proposed non-universal form of quantum computing.
It has often been considered a toy problem: Potentially proving quantum supremacy\footnote{The exact definition of quantum supremacy and its relation to the phrases `quantum advantage' or `quantum utility' are not the topic of this section and is assumed to be uncontroversial here.} over its circuit model counterparts, but effectively without practical applications~\cite{Aaronson2011-tm}. 
However, there is an increasing school of thought that boson sampling does have the potential to bring quantum advantage in a number of real world, computationally challenging problems \cite{Bromley2020-uc}. 

Aside from quantum advantage and quantum supremacy, there are other measures of utility that may motivate the development of photonic quantum computers.  
There is growing concern over the environmental impact of supercomputers and large data centres and quantum computers may be more energy efficient \cite{Sood2024-sp,Arora2024-tg}.
This is especially likely to be true of boson samplers as photons are stable at room temperature and do not require cryogenic cooling like superconducting qubits. 

Here we use a boson sampling simulator to
solve a dominating set problem with a surveillance use case on the ORCA TBI simulator, as a benchmark for the use of these devices against classical algorithms. The use of simulation allows us to consider the optimal performance of the computing paradigm without having to factor in noise, errors and scaling limitations. 
Future work should extend this study to the real hardware to consider both optimisation success and the timing compared to both the simulator and the classical algorithms. 

\section{Background}
\subsection{Boson Sampling Theory} \label{sec:BSTheory}

Boson sampling involves sampling from a distribution of identical bosons (typically photons) that have been through a linear interferometer.

A linear interferometer contains a number of beamsplitters, optical components used to split a beam of light into transmitted and reflected components. 
When a photon reaches a beamsplitter, it can either be reflected or transmitted, traversing a different path through the interferometer, 
with a probability depending on the \textit{angle} of the beamsplitter.
Photon detectors are placed at the end of each possible paths; the number of possible end locations is the number of \textit{modes} in the system. 
The count from an end location is the occupation of that particular mode. 
Two key parameters in practical boson sampling setups are the number of input photons, $N$, and the number of modes in the system, $M$. 
The total occupation over all the modes should be equal to the number of input photons, however these are often thresholded to given a binary readout of occupied and unoccupied modes.

\begin{figure}
    \centering
    \includegraphics[width=\textwidth]{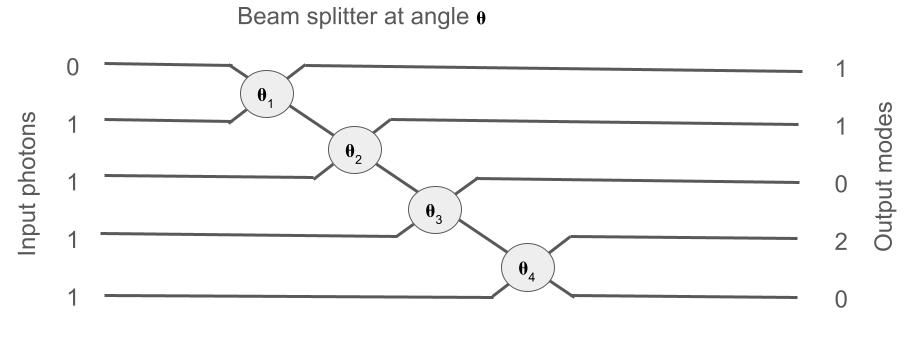}
    \caption{A simple boson sampling set up with 4 input photons, 5 output modes and 4 variable beam splitters each with angle $\theta_i$.}
    \label{fig:BosonS}
\end{figure}

When a photon hits a beamsplitter, the probability of reflection or transmission is controlled by a parameter called the \textit{angle} of the beamsplitter. 
With multiple indistinguishable photons and multiple beamsplitters, there are a number of ways that each possible output state could be obtained; calculating the expected distribution in advance is a  computationally expensive problem, equivalent to calculating the permanent\footnote{The permanent of a matrix is similar to the determinant in that it is a polynomial of the matrix coefficients \cite{Marcus1965-ic},
but unlike the determinant, there is no known efficient classical algorithm to compute it.} of the matrix that characterises the interferometer~\cite{Gard2015-cs}.
Figure \ref{fig:BosonS} shows an example small interferometer. 

The Hong-Ou-Mandel effect describes the non-classical outcome of two indistinguishable photons coincident on a beam splitter. 
It states that the two photons always exit in the same output mode, and never take different paths \cite{Hong1987-jq}. 
The outcomes where the photons would end in opposite modes destructively interfere, and therefore have a probability of 0. 
This occurs only if the two photons are identical and coincident, which stresses the importance of  controlled photon generation and tightly controlled distances between components. 
This has also been generalised to multi-photon interference \cite{Tillmann2015-gp}. 

The Hong-Ou-Mandel effect is the key element in making boson-sampling a quantum process because it relies on the quantised photonic nature of light.
The resulting path of the photonic pair is unknown until detection; we can think of the photons taking a superposition of the possible paths. 
Once detected, the superposition collapses and the path taken by the photon is known through the mode it was detected in.

Calculating the amplitude of all terms in the output state is \#P-complete\footnote{The \#P complexity class covers the function problems associated with the decision problems within the NP (non-deterministic polynomial time) class. As opposed to simply stating whether a solution exists, the question is how many solutions exist. The \#P class is mostly used in the discussion around approximation algorithms and combinatorics \cite{UnknownUnknown-gz}.}, and there are an exponential number of terms in the output \cite{Aaronson2011-tm}. Even approximating these amplitudes up to a multiplicative error is \#P hard. 
It is challenging to classically simulate boson sampling for more than 10s of photons.

Photonics is a promising medium for quantum computation: photons are stable at room temperature, so do not require expensive cryogenics. 
They do not react strongly with the environment.
They move at light speed, 
which presents both benefits and challenges when working with photonic qubits. 
The high speed can make operations difficult to control, and it requires high precision to get indistinguishable photons in terms of timings. 

A large source of error in photonic systems are `lossy components'.  Theoretical beamsplitters as lossless; there is no chance of photon absorption
In reality, beamsplitters can absorb photons, resulting in fewer output than input photons \cite{Barnett1998-vj}. 
Another source of error is `dark counts', where a photon detector  registers a detection even though no photon is present. This is likely to be a random error, and its likelihood of occurring should be determined by the manufacturer of the device. It is also possible for the performance of detectors to deteriorate over time. Therefore, detectors should be calibrated frequently, to ensure the error rate of dark counts is known and of an acceptable level. 

In simulation and in theory, beam splitter angles can be set to arbitrary precision, but in experimental set up, there is a limit to the achievable precision. 
There may also be systematic errors in the setting of beam splitter angles, such as a constant offset.
This can be tested and calibrated in small setups.

The photons needs to be indistinguishable, both temporally and in frequency. 
Any temporal mismatch will cause errors. 
Small temporal displacement of the wavepacket will manifest as dephasing, but when this becomes larger, it may become ambiguous which time-bin the photon should be in. 
Temporal displacement can be caused by errors in the path lengths, implemented by fibre loops, or by jitter in the photon source. 
Errors caused by fibre loop path lengths could potentially be replaced by a quantum memory or free space delay line. 

\subsection{ORCA PT Series}

\begin{figure}[tp]
    \centering
    \includegraphics[width=0.7\textwidth]{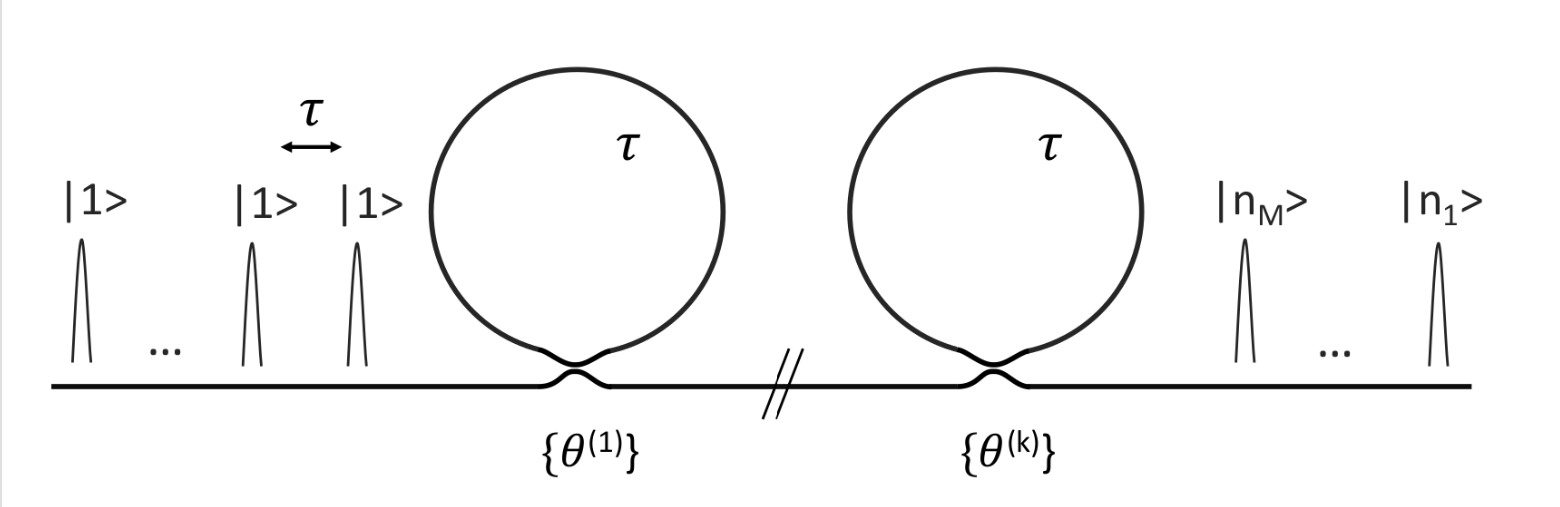}
    \caption{Diagram showing time-bin interferometry. Taken from \cite{Bradler2021-xa}}
    \label{fig:PT-Series}
\end{figure}

ORCA computing have designed and built a boson sampler than uses a time-bin interferometer \cite{Bradler2021-xa} (unlike the spatial binning of Figure \ref{fig:BosonS}), see Figure~\ref{fig:PT-Series}. 
Time-binning uses only a single photon source and a single detector.
Photons are inserted into the interferometer at regular time intervals;  beam splitters control the access to delay loops.
The photon modes are determined by the time-bins over which the detector may or may not receive a photon \cite{Motes2014-xj,He2017-pc}. 

To translate the spatial boson sampler into a time-bin interferometer,  consider the input photons entering a single path at regular time intervals, called the characteristic time-step of the interferometer. Where there is a photon on a spatial input path, there is a photon entering the time-bin interferometer at that time-step.  
When a photon reaches a beamsplitter, there is a probability (dependent on beamsplitter angle) that the photon is transmitted and continues on its path, and a corresponding probability the photon is reflected and enters a loop of fibre, returning to the same beamsplitter at a multiple of the characteristic time-step later.
Photons can interact at a beamsplitter when an early photon in the sequence has been suitably delayed within a loop such that its arrival is coincident with a later photon. 
The lengths of the loops are precisely defined such that the photons can be coincident on the beamsplitter.
The output modes of the time-bin interferometer are determined by a single photon detector at the output recording the number of photons that arrive at each time step.
Photons arrive at the detectors at time steps corresponding to the characteristic length of the input sequence. 
As with the spatial form, the photons need to be indistinguishable: when they  interact at the beam splitters it is not possible to trace which output mode corresponds to the input mode (at least in cases with multiple loops).

Using time-bin sampling rather than spatial bins allows the interferometer to be compact. 
Optical fibres and other telecommunication components can be purchased off-the-shelf. 
These factors have allowed ORCA Computing to build their first prototype, the PT-1, in a standard server rack. 
The PT-1 is a simple test of the hardware components and software processes, and is formed of a single loop and beam splitter. 
As soon as a photon is detected at the end of the interferometer, the user knows which path the photon has taken, which collapses the quantum state.
For a single loop, then, there are no `quantum' effects, since the presence (or lack) of a detection collapses the state at every time step. 
This limits the capability of this test-bed device in providing quantum utility, but it is useful for experimenting with the kinds of problems that these devices may be able to solve, and learning more about the parameters  involved.
Future iterations of the technology include multiple loops, which is where the quantum effects are expected, taking the capability beyond that which is classically simulatable. Multiple loops and multiple beam splitters is also likely to increase the errors in the system.

ORCA Computing have developed and released a Python-based Software Development Kit (SDK), which leverages the widely-used PyTorch package for machine learning applications \cite{Bradler2021-xa}.
The SDK includes a simulation capability and functions for solving problems on either the inbuilt simulator or on a real device.

\section{Methodology}
\subsection{Benchmarking Design}
Here we use the SDK to benchmark a particular use case.
The design phase is guided by the methodology laid out by the authors \cite{Park2024-dl} and other work on the principles and key steps of quantum benchmarking \cite{Eisert2020-fz,Wang2022-ud}. 
It first clarifies the purpose of the benchmark, then defines success and test selection based on the key principles of generalisability, robustness and expressivity \cite{Park2024-dl}. 
The key factors influencing the  choice of benchmark problem are:
\begin{enumerate}
    \item As our first exploration in boson sampling, to understand how it works and what it was designed to do.
    \item Quantifying to what extent the device does what it has been designed to do, in this case, solve optimisation problems. 
    \item Binary variables are native for the solver
    \item Quadratic Unconstrained Binary Optimisation (QUBO) problems are common benchmark problems and have been cited as a use case for boson solvers
    \item  A problem with a generalisable structure with potential applications
    \item A problem that is easily scalable and can be tested with a number of examples of different sizes
    \item A problem with known bounds for classical and/or alternative quantum algorithms. 
\end{enumerate}

A problem that fits the criteria listed and chosen for the benchmark assessment is the NP-hard \textit{Surveillance Coverage Problem}.
In its simplest form, this reduces to the problem of determining a minimum dominating set of a graph:
given a graph $G=(V,E)$, find a subset $V'$ of $V$ such that every node in $V$ is adjacent, via an edge in $E$, to at least one node in  $V'$ \cite{Grandoni2006-mz}. 

The nodes $V$ of the graph may model points of interest (POI) over a large region, where an edge represents line of sight between POIs, which could be affected by distance and obstacles.
The minimal dominating set represents the minimal number and optimal positions of surveillance equipment 
such that every POI is under surveillance. 

This fits our requirements, as it can be extended in size and complexity by adding more nodes or assigning some targets more value than others.
It is an optimisation function with a well defined cost function that fits in with ORCA PT series designed use cases. 
As a well-studied problem in graph theory, we have bounds for the various classical algorithms that we can compare against. 

\subsection{Solving the dominating set problem}
Consider a set $V'$ defined by a bit string $\bm{x} =\{x_0, x_1, ..., x_{n-1}\}$, where $n$ is the number of vertices in the graph $G$, and $x_i = 1$ if the vertex is in the set and $0$ otherwise. (Hence the set $V$ of all vertices in $G$ is the all-ones bitstring.)
The neighbouring vertices connected to vertex $i$ by some edge in the graph are denoted by the set $N(i)$.
The cost function $F$ to be minimised is:

\begin{equation} \label{eq:F}
    F(\bm{x}) = \sum_{i=0}^{n-1} \left( x_i + A\,P_i \right)
\end{equation}
where the $x_i$ term gives the size of the set,
$A$ is a scaling factor (typically $A=2$),
and $P_i$ is a penalty term:
\begin{equation} \label{eq:P}
    P_i = 
    \begin{cases}
    0,& \text{if } (x_i + \sum_{j \in N(i)}x_j) -1 \ge 0\\
    1,& \text{otherwise}
    \end{cases}
\end{equation}
The penalty $P_i$ term checks whether the vertex $i$ is either in the set $\bm{x}$, or is a neighbour of some vertex in the set. 
For every vertex in the graph that passes this check,
there is no penalty;
for every vertex that fails,
a penalty of $1$ is added.
If no penalty terms are added, the set in question is a dominating set and the result of the cost function is the size of the set.
We refer to the result of the cost function as the \textit{state energy} of that bit string.\footnote{
Here we are using the  definition in which a node can dominate itself. 
This models a surveillance coverage problem in which a sensor at a particular POI monitors its own location as well as other POIs within its line of sight.
For problems where this is not the case and a node cannot dominate itself, we have a strongly-dominating set or a total dominating set.
For such a case the cost function would not have the $x_i$ term in equation \ref{eq:P}. This then requires that each vertex be a neighbour of a member of the set without considering whether it is part of the set itself.
}


The minimisation is completed via a variational quantum algorithm shown in algorithm \ref{alg:VQA2},
which involves a learning process to find the problem-specific angles $\theta_i$. 
The input cost function, $F(x)$,  is derived from the graph and is calculated via equations \ref{eq:F} and \ref{eq:P}. The input state used in this algorithm is a state with number of input modes equal to the number of vertices in the graph, $n$. 
The occupancy of the input modes alternates between 1 and 0 photons.
For example when $n=6$, the input state is $\ket{101010}$.
This is the default input state provided in the ORCA SDK and advice from the ORCA Computing team \footnote{Personal communication} suggests that they believe it to provide the most non-deterministic behaviour and therefore more efficiently explore the search space. 
As far as we aware, there are no published results on the effect that changing the input state has on the results.
There may be input states that are more suited to this (or any particular problem) but the state that maximises output entropy is likely to be a generically good choice.
This is an area of active research at ORCA Computing and an open question for further research. 

\begin{algorithm}[tp]
\caption{Variational Quantum Algorithm}\label{alg:VQA2}
\textbf{Inputs:} cost function $F(\bm{x})$, input state $\ket{10} \otimes \frac{n}{2}$ \newline
\textbf{Outputs:} Optimum state $x_{min}$, Minimum energy $E_{min}$
\begin{algorithmic}[1]
\State Initialise interferometer parameters to random values $\bm{\theta} = \{\theta_i\}$
\State Initialise flipping probabilities to random values $\bm{p} = \{p_i\}$
\Repeat 
    \Repeat
        \State Pass input state through the interferometer and measure output photons in each mode $\ket{out}$
        \State Convert $\ket{out}$ to binary bit string via Threshold Mapping, $\bm{x}$  \label{algline:convert}
        \State Flip or hold bits in $\bm{x}$ depending on $\bm{p}$  \label{algline:flip}
        \State Calculate state energy, $E = F(\bm{x})$
        \Until $maxSamp$ is reached
    \State Use classical optimisation algorithm (SPSA) to update $\bm{\theta}$ and $\bm{p}$ \label{algline:classicalopt}
    \Until{$maxIter$ has been reached}
\State \textbf{Return} $b_{min}$, $E_{min}$
\end{algorithmic}
\end{algorithm}

Algorithm \ref{alg:VQA2}, line \ref{algline:convert}, 
the conversion from output state to a binary bit-string, 
is performed using the Threshold Mapping,
in which any mode that has at least one photon detected is given a value 1 and modes with no photon detected are given the value 0. 
Line \ref{algline:flip} describes the second part of the threshold mapping process, which involves probabilistically flipping the bits of the bit-string $\bm{x}$ to generate a new candidate set. 
The probabilities of flipping each bit are 
updated in each iteration of the training process.

The classical optimisation, line \ref{algline:classicalopt}, is Simultaneous Perturbation Stochastic Approximation (SPSA), which is often used to optimise multiple unknown parameters \cite{Spall1992-yh}, 
and is the standard optimisation algorithm in the ORCA SDK.    
It follows a similar optimisation rule to standard gradient descent, calculating the gradient for all of the parameters together using two function estimations per iteration.
SPSA has been shown to perform better when optimising noisy or error-prone functions \cite{Finck2011-le}: 
every parameter is shifted stochastically in each step even in the perfect case, making the whole process more robust to noise in the cost function \cite{Szava2023-wj}. 

We run this for a set number of iterations, $maxIter$, with a fixed learning rate (representing how much a parameter is changed per iteration) and number of samples, $maxSamp$. 
At the end of the training, the bit-string with the lowest energy is returned. This is classically checked for being a dominating set.

Test graphs are generated using the NetworkX python package and the function $fast\_gnp\_random\_graph$. This function takes in three parameters: 
$n$ is the number of nodes in the graph (here called the problem size);
$p$ is the probability for edge creation (graph density);
$s$ is the random seed. 
For each combination of $(n, p)$, different seeds are used to produce alternate graphs for repeat tests. 

The following values are recorded for each run:
\begin{enumerate}
    \item Is the solution a dominating set? (Y/N)
    \item Size of the found set (\# of nodes)
    \item Total time to train (for one run, no restarts, over fixed number of iterations)
    \item Time per update (total time divided by number of iterations)
    \item Iterations required to converge
    \item Time require to converge (the product of the two previous measures) 
\end{enumerate}

It is computationally expensive to verify whether a dominating set is  minimal, so we do not use this as a stand-alone measure. 
We use the size of the dominating sets found to compare between different implementations run on the same graph. 

The ORCA SDK does not provide any timing metrics natively, so these are built on top of the training algorithm. 
This inefficiency might add some time to the measure, but we assume  that it  negligible compared to the time to train. 
This assumption is monitored throughout the experiment. 

We define training to have converged when the minimum energy has not decreased over the previous 50 iterations. This method for testing convergence was deemed appropriate after some preliminary tests on example problems.

For the classical comparison, the measures also include success rate (does the algorithm find a dominating set), the size of the found set and the time taken to execute the algorithm. As for many NP-hard problems, there are approximate algorithms that are more efficient than exact algorithms, but have a non-zero probability of failure.  

The first classical algorithm is based on integer linear programming methods~\cite{Forrest2005-uh}. 
We use the python package PULP with the default solver. 

The second classical algorithm is a basic greedy algorithm.
It loops over vertices not yet in the set, adding the one with highest number of neighbours to the set \cite{Esfahanian2013-wt}. 
All of the neighbours are then marked as `visited'. 
The loop ends when all the vertices in the network have been visited. 
This algorithm never fails to find a dominating set, but it may be much larger than the minimal set.

The third classical algorithm is the approximation algorithm in the NetworkX python package, which runs in $\mathcal{O}(\# E)$ time.
It follows the $k$-centre algorithm of \cite{Vazirani2001-zj}, which uses the relationship between this problem and maximally independent sets.

These three classical methods are compared to the simulated ORCA PT-1 in terms of the size of the dominating set found and in the wall clock time to run. 
All the results in this paper have been produced with an Intel i7 processor with 64GB of RAM. 

\section{Results and Discussion}
\subsection{ORCA TBI Simulator}
\begin{figure}[tp]
    \centering
    \includegraphics[width=0.49\linewidth]{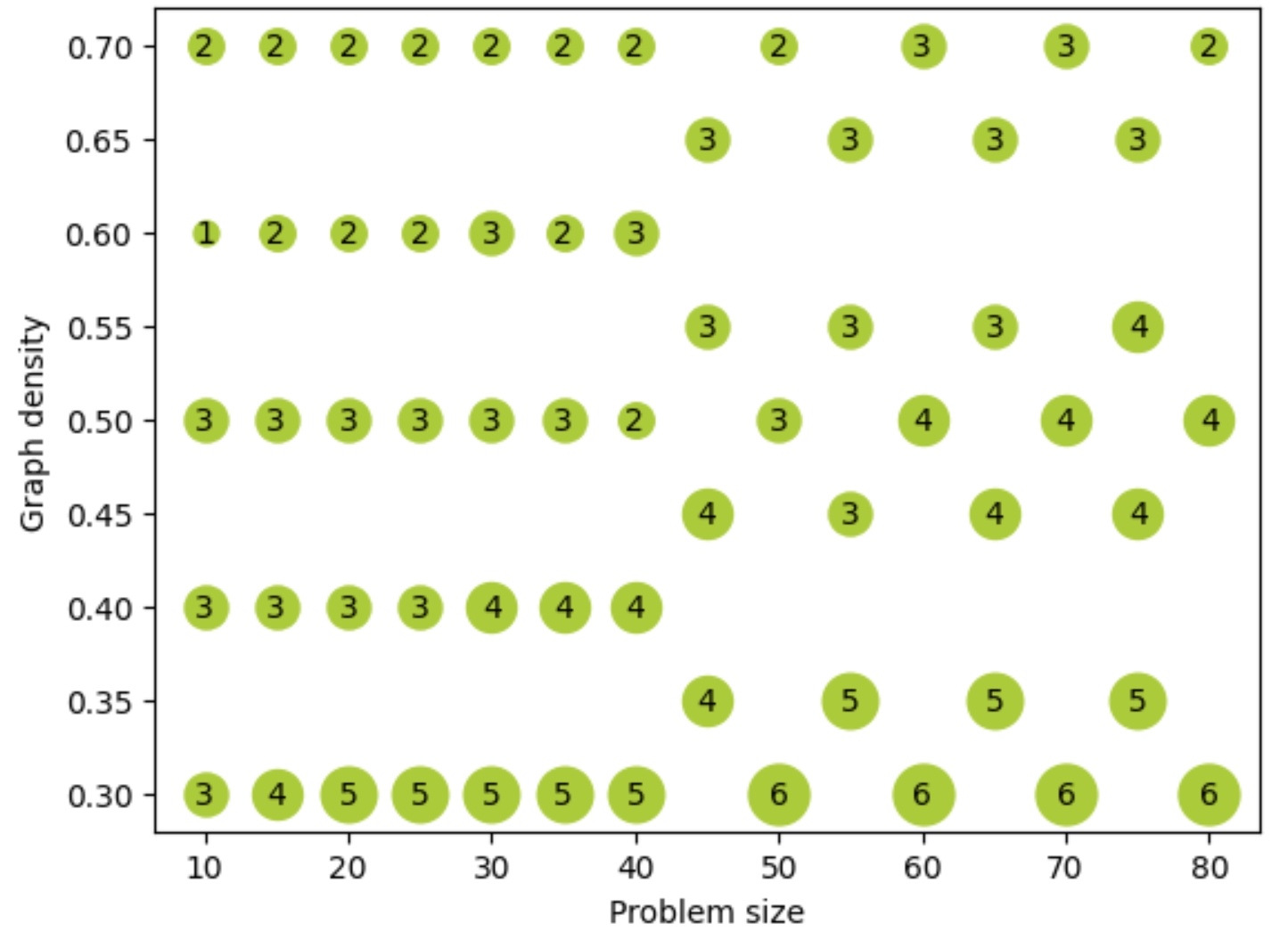}
    \caption{Summary of test runs on the TBI Simulator. Markers are sized and labelled with the size of the dominating set found.}
    \label{fig:TBISumm}
\end{figure}

This section covers the results from the Time-Bin Interferometer Simulator provided in the ORCA SDK. 
In this case, the TBI Simulator is  used to simulate a PT-1 type machine with a single loop and beam splitter.
Figure \ref{fig:TBISumm} shows a summary of the individual tests carried out. 
Each test has a unique graph associated with it that is characterised by the problem size $n$, number of vertices, and graph density $p$, the probability of an edge existing between any two vertices. 
Higher $n$ and lower $p$ values characterise harder problems. 
In figure \ref{fig:TBISumm} the markers represent that all of the tests performed successfully found a dominating set. 
The size of the markers indicate the size of the dominating set found by the TBI simulator. 
The measure of success here is whether alternative, classical, methods fail to find dominating sets smaller in size than the quantum simulator.

\begin{figure}[tp]
\centering
\begin{minipage}[t]{0.49\textwidth}
    \centering
\includegraphics[width=0.95\textwidth]{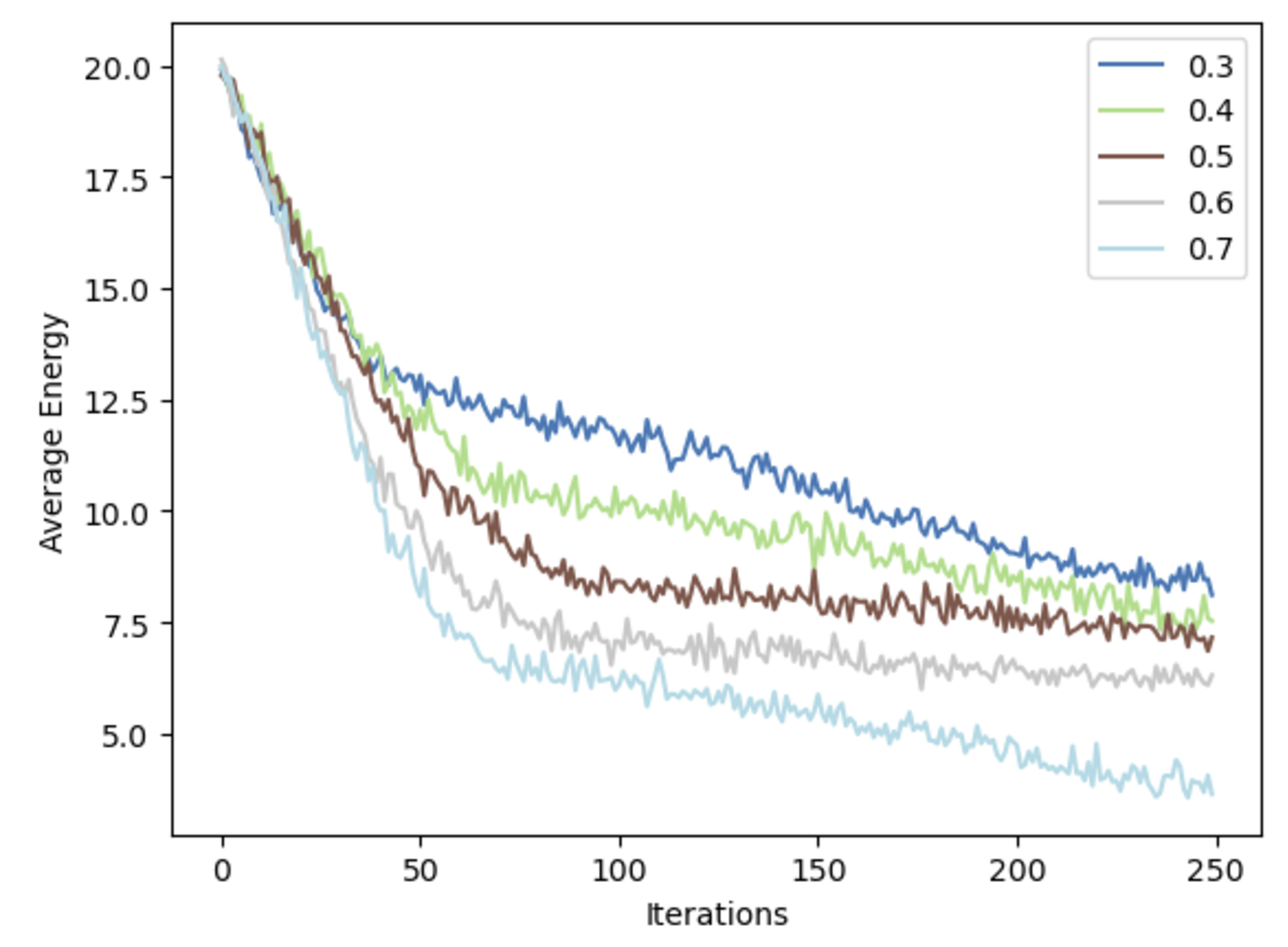}
\subcaption{Problem size fixed ($n=30$) with varying graph density ($p$).\\\vspace{1em}} \label{fig:N30}
\end{minipage}
\begin{minipage}[t]{0.49\textwidth}
    \centering
\includegraphics[width=0.95\linewidth]{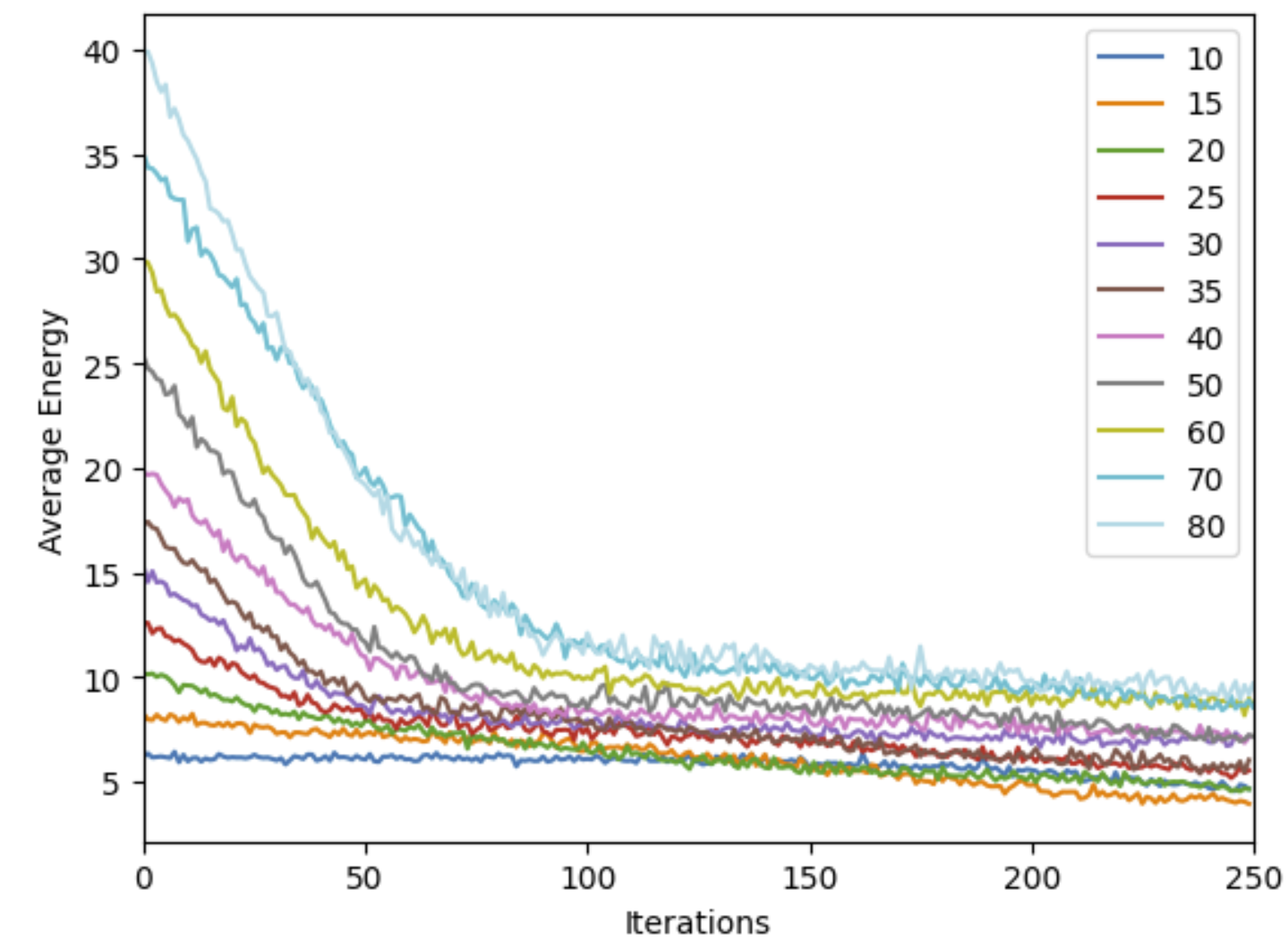}
\subcaption{Fixed $p=0.5$ for a range of $n$ values.} \label{fig:p05}
\end{minipage}
\caption{The average energy of the samples, as given by Equation \ref{eq:F}, at each step in the training process.} \label{fig:Logs}
\end{figure}

The TBI simulator outputs a log that documents the training process. Examples of these logs are shown in figure \ref{fig:Logs}.
Figure \ref{fig:TBISumm} shows that the size of the dominating set tends to decrease as graph density $p$ increases; this can also bee seen in figure \ref{fig:N30}.
This is expected because a higher number of edges (within graphs of the same size) should mean that fewer vertices are needed to dominate the graph. In the surveillance coverage problem, a higher density of connections might imply a flatter landscape where a POI has line of sight to many more POIs than in a rugged landscape. 

\begin{figure}
    \centering
    \includegraphics[width=0.6\linewidth]{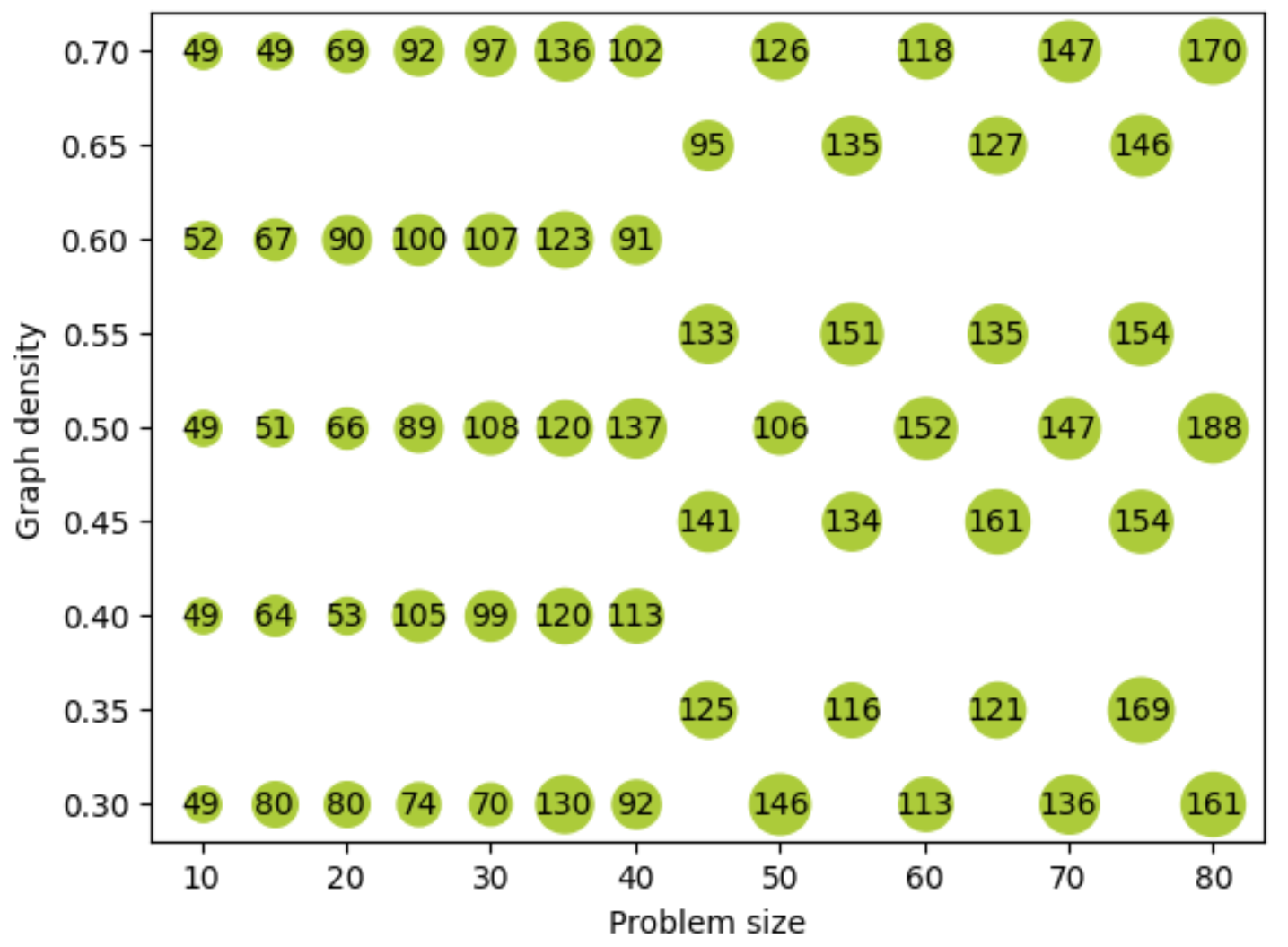}
    \caption{Summary of test runs on the TBI Simulators. Markers are sized and labelled with the number of iterations required for convergence according to the convergence test laid out in the methodology section.}
    \label{fig:Timing}
\end{figure}

The time taken to run training also scales with both problem size and graph density. 
Problem size defines the length of the bit-string, and graph density defines the probability that each possible edge will be included in the graph ($p=1$ produces a complete graph). 
Larger values extend the run time of the cost function. 
Figure \ref{fig:Timing} follows the same format as figure \ref{fig:TBISumm}, except now the markers are sized and labelled according to the number of iterations required for convergence. 
As stated in the methodology, convergence has occurred when the minimum energy has not decreased for 50 consecutive iterations. 
All of the tests here converge before the 250\textsuperscript{th} iteration. 

\subsection{Classical Method Comparison}
\begin{figure}
    \centering
    \includegraphics[width=0.49\linewidth]{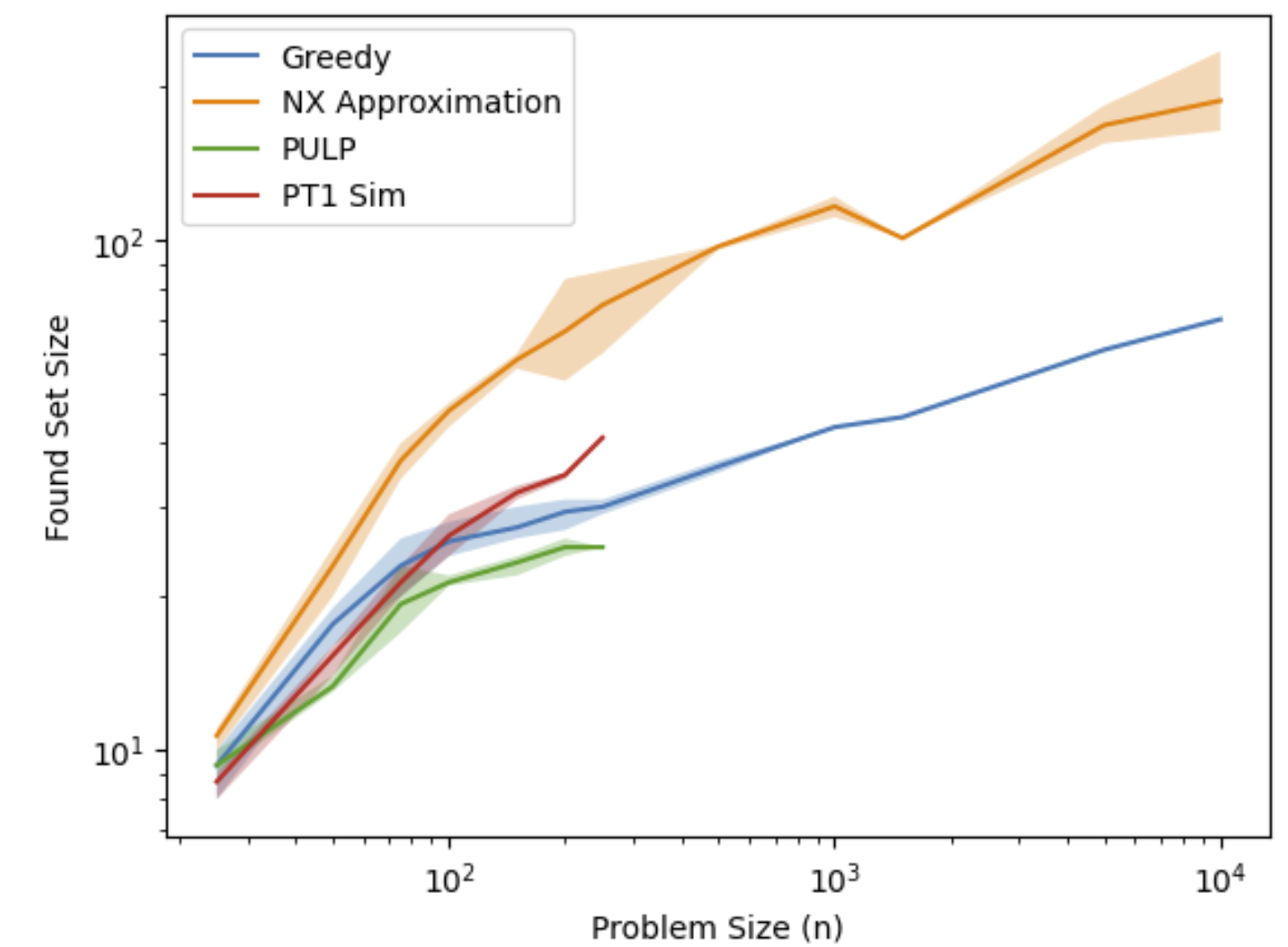}
    \includegraphics[width=0.49\linewidth]{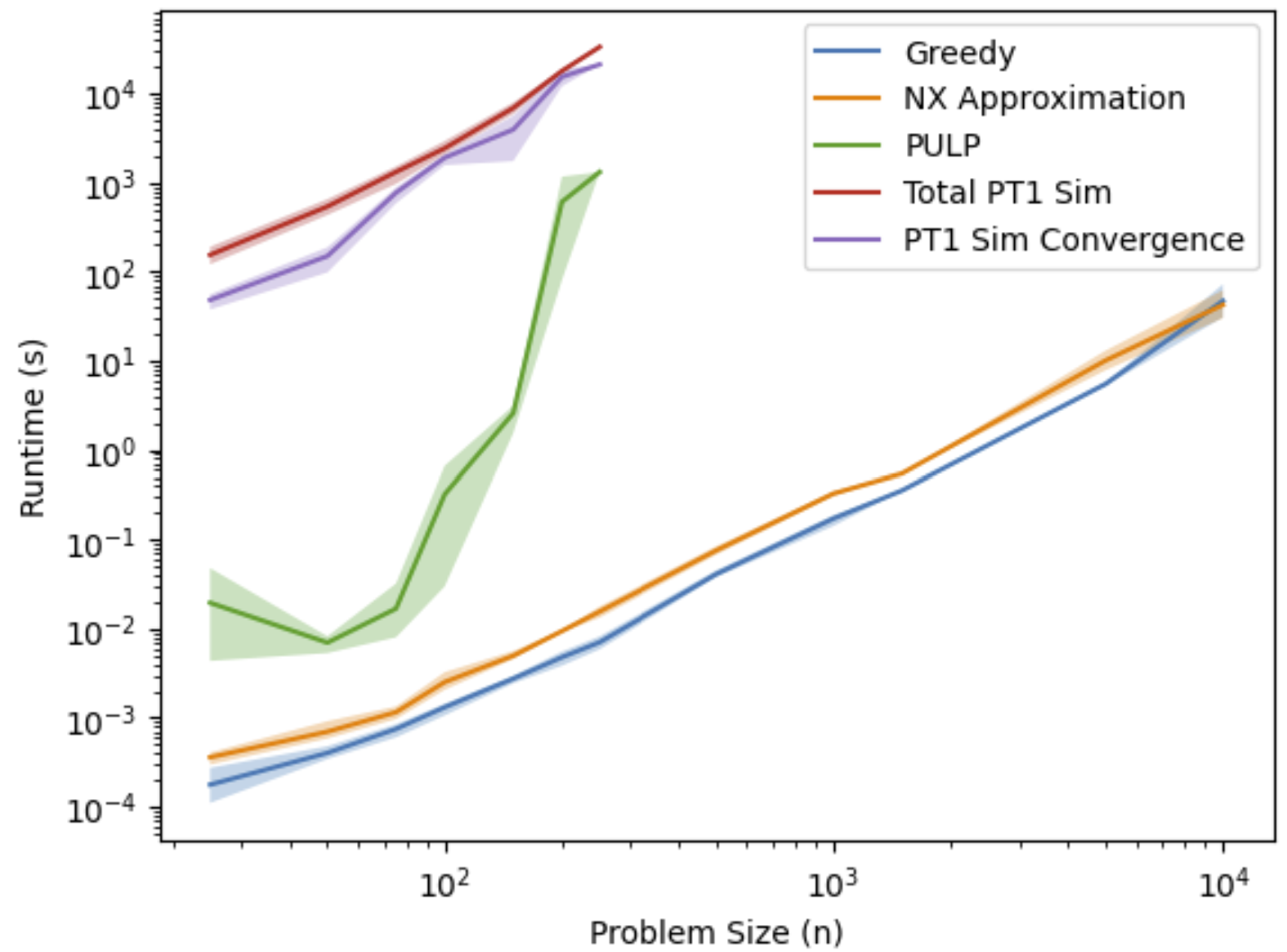}
    \caption{Comparisons between the three classical methods for solving the dominating set problem and the TBI Simulator. Note the log scaling. The lines represent the average value over three different graphs (of the same $n$ and $p$ values) and the shaded areas shows the spread of results. The left-hand graph shows the minimum dominating set size found; the right-hand graph shows the wall clock runtime. The convergence runtime is calculated to be the time at which the minimum energy has not decreased from the previous 50 iterations.}
    \label{fig:Comp1}
\end{figure}

Figure \ref{fig:Comp1} shows a comparison of the three classical methods used in this test against the TBI Simulator. 
The left graph is a comparison in terms of the size of dominating sets found;
the right graph compares the methods based on their wall clock runtime.
The runtime is measured just for performing the algorithm and does not include the time to save out the results. 
For all of the graphs in these tests, the graph density parameter is set to $p=0.05$.  

The Greedy and NetworkX algorithms both run in the approximate millisecond regime for the problem sizes tested here, and also show good scaling that would allow the solving of much larger problems in times less than the PULP solver, even at modest problem sizes. 
However, the NX Approximation algorithm consistently finds larger dominant sets than the other method. 
The Greedy algorithm finds dominant sets of an approximately equal size to that of the TBI Simulator.  
The PULP method typically finds the smallest dominant set out of all the methods tested here, however the time taken is approximately an order of magnitude larger than the other two classical methods for $n<100$ and scales increasingly badly beyond this. 
Neither the TBI Simulator or the PULP methods could produce results beyond $n>250$, due to prohibitive runtimes. 

Both the runtime and the minimum set size of the TBI simulation results are affected by the parameters chosen in the algorithm, such as the number of samples and the learning rate. 
The results shown here are not claimed to be an optimal setting, but may be indicative of likely performance results that could be achieved. 
The optimal parameters for the simulator will likely depend on the problem and its formulation, as well as the priorities of the user. 
In the case where a near-minimal dominating set is sufficient, the run time could be reduced by doing fewer samples or iterations.

\section{PT-2 Preliminary Results}
ORCA Computing have recently released the PT-2 series of Time-Bin Interferometers \cite{Unknown2024-dx} with two loops (two beam splitters) and upgraded photon generation. 
A single loop cannot exploit superposition, as either the photon is detected upon leaving the beam splitter or it is not detected and we are certain it remains in the loop. 
With two loops, there is ambiguity in the location of the photons and it is likely that the effects of superposition can be exploited to solve problems. 

In preparation for this new hardware, we perform some preliminary investigations using the TBI Simulator to emulate a PT-2 device, with the aim to compare these both to PT-1 simulation results described above, and to experimental results in the future. 

\begin{figure}[tp]
    \centering
    \includegraphics[width=0.49\linewidth]{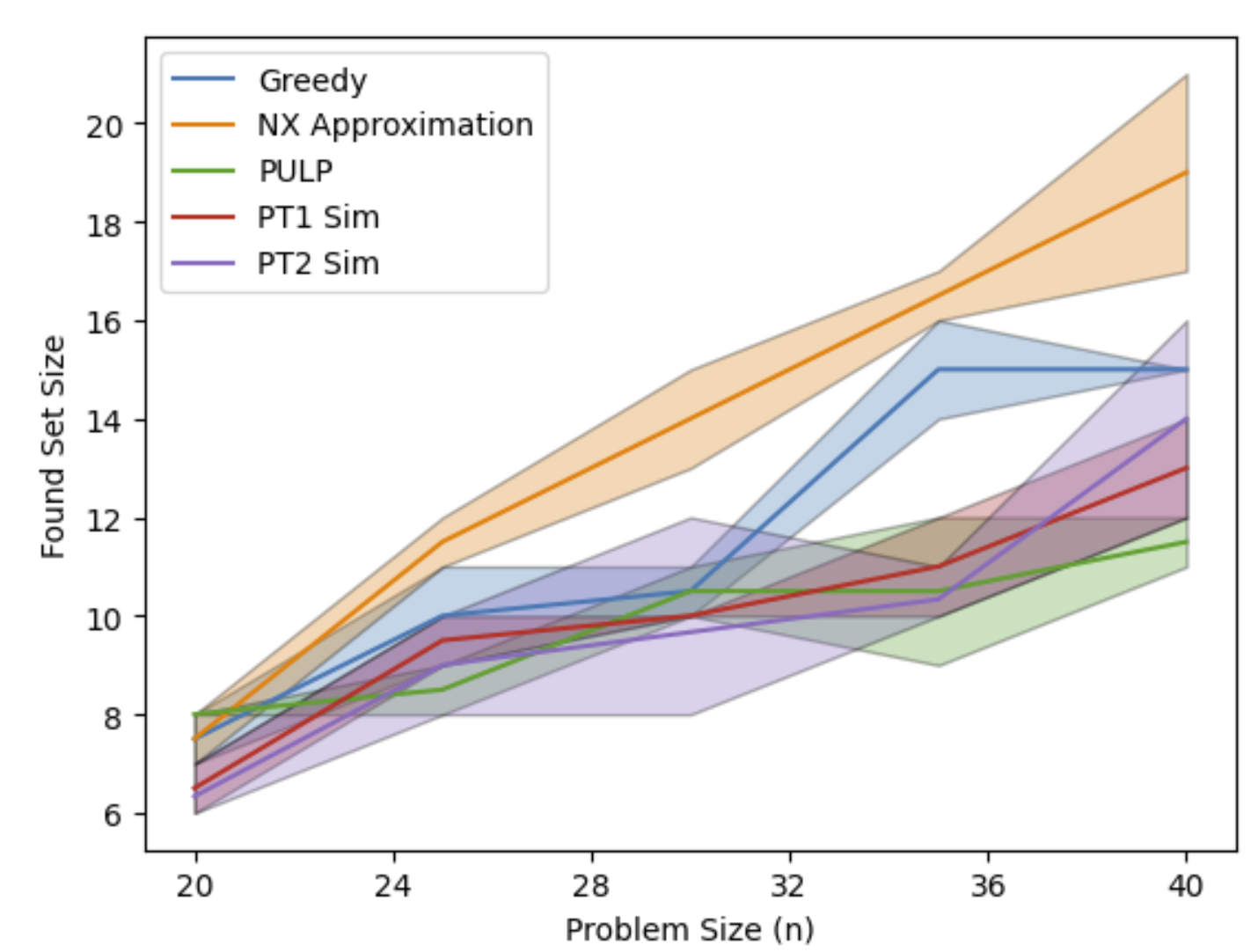}
    \includegraphics[width=0.49\linewidth]{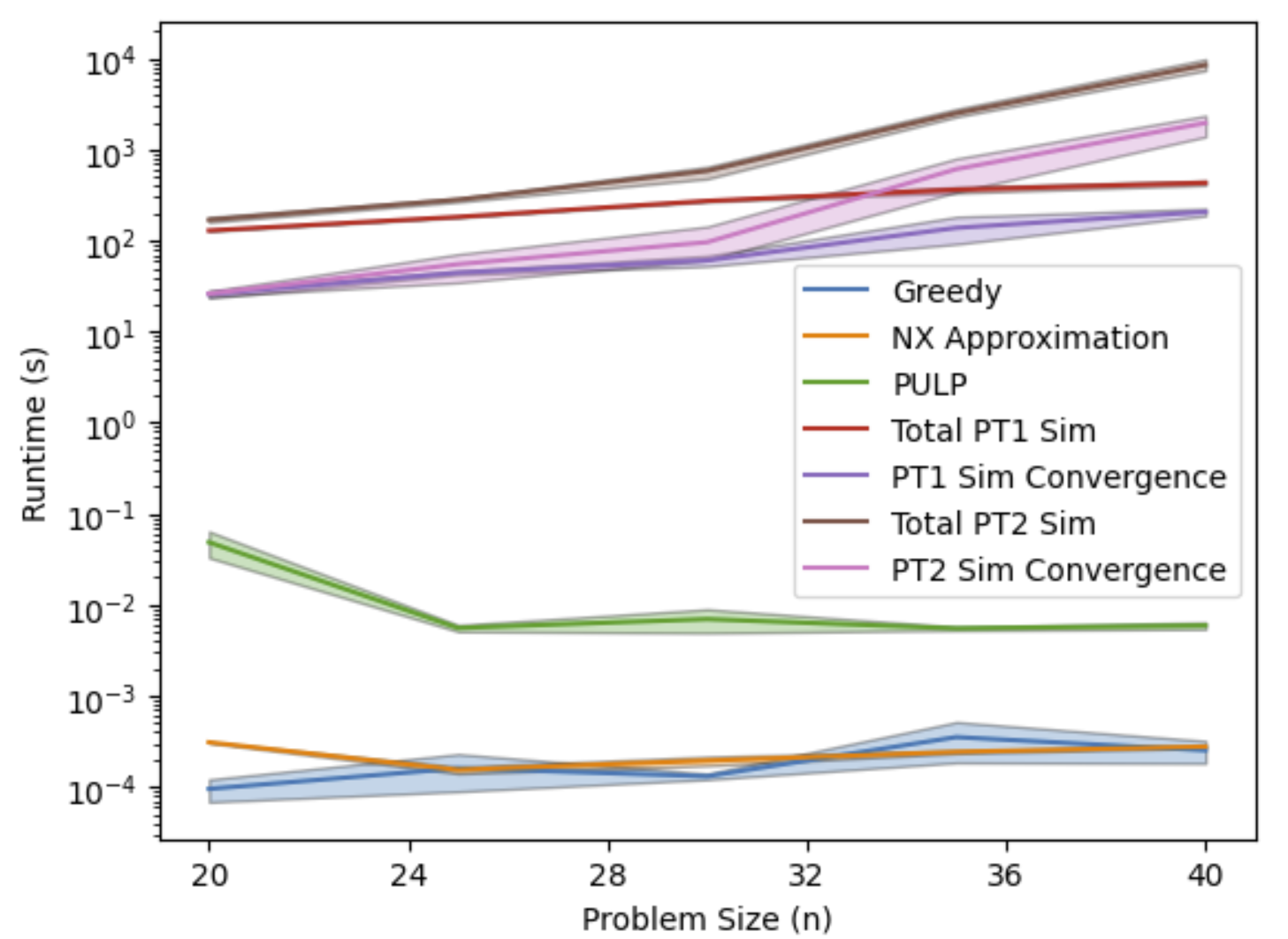}
    \caption{Comparisons between the three classical methods for solving the dominating set problem and the TBI Simulator in both PT- 1 and PT-2 configurations. Note the log scaling on the y-axes of the runtime graph only. The lines represent the average value over three different graphs (of the same $n$ and $p$ values) and the shaded areas shows the spread of results.}
    \label{fig:Comp2}
\end{figure}

The TBI simulator can be used to simulate time-bin interferometers with any number of loops of differing sizes. 
However, there are limitations on the size of systems that can be simulated.
%
The power of the computer the simulator is run on affects both the time the simulator takes to run and the maximum problem size that can be run on the simulator. 
Due to this, only problem sizes $n \le 40$ are able to be tested here. 
To compare these to the PT-1 results, Figure \ref{fig:Comp2} shows a zoom in of Figure \ref{fig:Comp1} with additional lines for the PT-2 simulation. 

The two-loop simulation shows comparable quality results to the single loop simulation, but the limitation on solvable problem sizes has allowed only a small sample size of results. 
The runtime is approximately a factor of 10 higher (for the problem size $\sim$ 40) than the single-loop scenario, but shows a similar scaling.  
In terms of the found set size, the two-loop scenario appears to slightly outperform the other methods, but does show considerable variability, making it difficult to conclude whether there is additional benefit in using the second loop. 
It could be the case that the additional benefit from using the second loop does not appear in the simulation case: if the quantum effects of superposition and entanglement that could allow a better exploration of the search space are only noticeable in larger problem sizes, they would be out of range of the simulation capability.
We aim to test this hypothesis in future work completed on the real ORCA PT-2 hardware.

\section{Experimental Conclusions and Future Work}
The results from the benchmarking experiment show that, at least as modest problem sizes, boson sampling is capable of solving a minimum dominating set problem. 
This problem can be interpreted in a problem specific context as a surveillance coverage problem, showing how boson sampling might be operationalised to provide benefit in such scenarios. 
In a less specific, more general sense, these results show that boson sampling has potential uses for a range of graph and optimisation problems. 

Although the results show some potential utility in the boson sampling methodology in simulation, the timing results show that these use cases might be niche and focused on areas in which minimising the size of the dominating set is really important and justifies the extra time that would be required to optimise the parameters and run the algorithms compared to the faster classical algorithms. 
Although we chose the problem to be representative of graph problems in general, we do not know whether the results seen here generalise to other problems. 
We might expect the real device to have much faster runtimes that would expand the potential use cases, but this will be tested in future work.

Limitations of this study are that it considers only relatively simple time-bin interferometer set ups with one or two loops.
ORCA have a roadmap for further devices that use multiple loops of various lengths; the current emulator is capable of simulating these set ups. 
A more complex interferometer, by providing more parameterised components, should be able to solve more complex problems in fewer iterations \cite{Bradler2021-xa}.
Understanding how the solution quality and runtime scale with the depth and complexity of the interferometer circuit will provide more information on the potential for boson sampling to perform comparatively with classical algorithms \cite{Slysz2024-hs}. 
This is currently an open question and a valuable option for further research.

\begin{credits}
\subsubsection{\ackname} The authors acknowledge Defence Science Technical Laboratory (Dstl) who are funding this research. 
The contents include material subject to © Crown copyright (2025), Dstl. This information is licensed under the Open Government Licence v3.0. To view this licence, visit https://www.nationalarchives.gov.uk/doc/open-government-licence/.  
Where we have identified any third party copyright information you will need to obtain permission from the copyright holders concerned. Any enquiries regarding this publication should be sent to: centralenquiries@dstl.gov.uk.

\subsubsection{\discintname}
The authors have no competing interests.
\end{credits}
%
%
%
\bibliographystyle{splncs04}
\bibliography{paperpile}

\begin{thebibliography}{10}
\providecommand{\url}[1]{\texttt{#1}}
\providecommand{\urlprefix}{URL }
\providecommand{\doi}[1]{https://doi.org/#1}

\bibitem{UnknownUnknown-gz}
{P} complexity class: {NP}, sharp {P} \& 2 sat. \url{https://www.studysmarter.co.uk/explanations/computer-science/theory-of-computation/p-complexity-class/}, accessed: 2024-8-22

\bibitem{Unknown2024-dx}
{ORCA} computing unveils the {PT}-2: Delivering quantum-enhanced generative {AI} capabilities. \url{https://orcacomputing.com/orca-computing-unveils-the-pt-2-delivering-quantum-enhanced-generative-ai-capabilities/} (Oct 2024), accessed: 2025-4-3

\bibitem{Aaronson2011-tm}
Aaronson, S., Arkhipov, A.: The computational complexity of linear optics. In: Proceedings of the forty-third annual ACM symposium on Theory of computing. pp. 333--342. STOC '11, Association for Computing Machinery (2011)

\bibitem{Arora2024-tg}
Arora, N., Kumar, P.: Sustainable quantum computing: Opportunities and challenges of benchmarking carbon in the quantum computing lifecycle. arxiv:2408.05679 [quant ph]  (2024)

\bibitem{Barnett1998-vj}
Barnett, S.M., Jeffers, J., Gatti, A., Loudon, R.: Quantum optics of lossy beam splitters. Phys. Rev. A  \textbf{57}(3), ~2134 (1998)

\bibitem{Bradler2021-xa}
Bradler, K., Wallner, H.: Certain properties and applications of shallow bosonic circuits. arXiv:2112.09766 [quant-ph]  (2021)

\bibitem{Bromley2020-uc}
Bromley, T.R., Arrazola, J.M., Jahangiri, S., Izaac, J., Quesada, N., Gran, A.D., Schuld, M., Swinarton, J., Zabaneh, Z., Killoran, N.: Applications of near-term photonic quantum computers: software and algorithms. Quantum Sci. Technol.  \textbf{5}(3),  034010 (2020)

\bibitem{Eisert2020-fz}
Eisert, J., Hangleiter, D., Walk, N., Roth, I., Markham, D., Parekh, R., Chabaud, U., Kashefi, E.: Quantum certification and benchmarking. Nat. Rev. Phys.  \textbf{2}(7),  382--390 (2020)

\bibitem{Esfahanian2013-wt}
Esfahanian, H.: Connectivity algorithms. Topics in structural graph theory pp. 268--281 (2013)

\bibitem{Finck2011-le}
Finck, S., Beyer, H.G., Melkozerov, A.: Noisy optimization: a theoretical strategy comparison of {ES}, {EGS}, {SPSA} \& {IF} on the noisy sphere. In: Proceedings of the 13th annual conference on Genetic and evolutionary computation. pp. 813--820. ACM (2011)

\bibitem{Forrest2005-uh}
Forrest, J., Lougee-Heimer, R.: {CBC} user guide. In: Emerging Theory, Methods, and Applications, pp. 257--277. INFORMS (2005)

\bibitem{Gard2015-cs}
Gard, B.T., Motes, K.R., Olson, J.P., Rohde, P.P., Dowling, J.P.: An introduction to boson-sampling. In: From Atomic to Mesoscale, pp. 167--192. World Scientific Publishing (2015)

\bibitem{Grandoni2006-mz}
Grandoni, F.: A note on the complexity of minimum dominating set. J. Discrete Algorithms (Amst.)  \textbf{4}(2),  209--214 (2006)

\bibitem{He2017-pc}
He, Y., Ding, X., Su, Z.E., Huang, H.L., Qin, J., Wang, C., Unsleber, S., Chen, C., Wang, H., He, Y.M., Wang, X.L., Zhang, W.J., Chen, S.J., Schneider, C., Kamp, M., You, L.X., Wang, Z., Höfling, S., Lu, C.Y., Pan, J.W.: Time-bin-encoded boson sampling with a single-photon device. Phys. Rev. Lett.  \textbf{118}(19),  190501 (2017)

\bibitem{Hong1987-jq}
Hong, C.K., Ou, Z.Y., Mandel, L.: Measurement of subpicosecond time intervals between two photons by interference. Phys. Rev. Lett.  \textbf{59}(18),  2044--2046 (1987)

\bibitem{Marcus1965-ic}
Marcus, M., Minc, H.: Permanents. Am. Math. Mon.  \textbf{72}(6),  577--591 (1965)

\bibitem{Motes2014-xj}
Motes, K.R., Gilchrist, A., Dowling, J.P., Rohde, P.P.: Scalable boson sampling with time-bin encoding using a loop-based architecture. Phys. Rev. Lett.  \textbf{113}(12),  120501 (2014)

\bibitem{Park2024-dl}
Park, J., Stepney, S., D'Amico, I.: A methodology for comparing and benchmarking quantum devices. In: UCNC 2024, Pohang, South Korea, June 2024. LNCS, vol. 14776, pp. 28--42. Springer Nature Switzerland (2024)

\bibitem{Slysz2024-hs}
Slysz, M., Kurowski, K., Waligóra, G.: Solving combinatorial optimization problems on a photonic quantum computer. arXiv:2409.13781 [quant-ph]  (2024)

\bibitem{Sood2024-sp}
Sood, V., Chauhan, R.P.: Quantum computing: Impact on energy efficiency and sustainability. Expert Syst. Appl.  \textbf{255}(124401),  124401 (2024)

\bibitem{Spall1992-yh}
Spall, J.C.: Multivariate stochastic approximation using a simultaneous perturbation gradient approximation. IEEE Trans. Automat. Contr.  \textbf{37}(3),  332--341 (1992)

\bibitem{Szava2023-wj}
Szava, A., Wierichs, D.: Optimization using {SPSA}. \url{https://pennylane.ai/qml/demos/tutorial\_spsa} (Mar 2023), accessed: 2024-11-28

\bibitem{Tillmann2015-gp}
Tillmann, M., Tan, S.H., Stoeckl, S.E., Sanders, B.C., de~Guise, H., Heilmann, R., Nolte, S., Szameit, A., Walther, P.: Generalized multiphoton quantum interference. Phys. Rev. X  \textbf{5}(4),  041015 (2015)

\bibitem{Vazirani2001-zj}
Vazirani, V.V.: Approximation Algorithms. Springer, 1 edn. (Jul 2001)

\bibitem{Wang2022-ud}
Wang, J., Guo, G., Shan, Z.: {SoK}: Benchmarking the performance of a quantum computer. Entropy  \textbf{24}(10), ~1467 (2022)

\end{thebibliography}

\end{document}